\documentclass[12pt]{article}
\usepackage{oldgerm}
\usepackage{yfonts}
\usepackage{euler}
\usepackage{graphics}
\usepackage{bm}
\usepackage{graphicx}
\usepackage{amsmath}
\usepackage{amssymb}
\usepackage{amstext}
\usepackage{amscd}
\usepackage{amsfonts}
\usepackage{appendix}
\usepackage{wasysym}
\usepackage{epstopdf}
\usepackage{color}
\DeclareMathAlphabet{\EuFrak}{U}{euf}{m}{n}
\DeclareMathAlphabet{\EuScript}{U}{eus}{m}{n}

\newcommand{\nd}{\noindent}

\title{{\bf The Tachyon Propagator}
\thanks{\it{This work was partially supported by Consejo Nacional de
Investigaciones Cient\'{\i}ficas and Comisi\'{o}n de Investigaciones
Cient\'{\i}ficas de la Pcia. de Buenos Aires; Argentina.}}}
\author{{D. G. Barci, C. G. Bollini, M.C.Rocca}, \\
\small{Departamento de F\'{\i}sica,
Universidad Nacional de La Plata,}\\
\small{C.C 67, 1900, La Plata, Argentina}}

\date{March 3, 1992}

\begin{document}

\maketitle

\vspace{-5mm}

\begin{abstract}

Following the canonical quantization procedure for a tachyon field,
the usual Hamiltonian and the creation and annihilation operators are obtained. The
observation that the mass hyperboloid $p^2-m^2=0$ is one-sheeted, as opposed to
the case of bradyons where $p^2+m^2=0$ is two-sheeted, leads to the construction of
a base which is unbounded for negative as well as for positive energies. There is a
zero-energy eigenfunction from which all other states can be constructed by
repeated application of decreasing or increasing operators; within this Fock space
the vacuum expectation value of the chronological product of field operators is
shown to coincide with Cauchy's principal-value Green's function.\\
\nd

\end{abstract}

\newpage

\section{Introduction}

Tachyon field excitations were considered in the literature every now and then.
For the classical theory of tachyons and a corresponding basic bibliography one can
consult ref. [1]. Also we think it interesting to mention that the classical radiation of
an accelerated charge can be interpreted as a Cerenkov radiation of the tachyonic
component of the movement [2].
Tachyonic field excitations appear for example in bosonic-string theories and the
problems related with unitarity were examined by Jacobson, Tsamis and Woodard [3]
and references therein. They appear also in higher-order Lagrangians, particularly in
those related with supersymmetry in higher dimensions [4].
There are several excellent studies about the quantum theory of tachyons.
For a description of the properties that tachyons would have we mention the
work by Feinberg[5] and that of Dhar and Sudarshan[6], where the sphere
k < m is removed when the field is quantized. The work by Kamoi and Kamefuchi [7],
where a review is given of different quantization methods, is also important.
It was shown in ref. [3] that it is practically impossible to construct a unitary
S-matrix for the tachyon field. This result implies that tachyons cannot be allowed to
appear in free asymptotic states. For this reason the tachyonic field should be
considered more as an auxiliary concept than as a real entity.
In this paper we do not intend to propose any model for tachyonic interaction. Our
aim is to show that, if a tachyon field appears at the Lagrangian level, then its
quantization leads to a propagator which is half-advanced plus half-retarded.
Furthermore, for methodological reasons, as well as for the sake of clearness we
decided to divide the work into two parts, each with peculiarities of its own. The first
part (this note) deals with the region of k-space where $k>m$. In a forthcoming second
part we are going to consider the sphere $k<m$.
We would like to point out that when the field is forbidden to appear in asymptotic
states, the only trace it leaves in the S-matrix is the propagator (unlike the case of
bradyons where the field is attached to the external legs).
The Wheeler propagator, (half-advanced plus half-retarded) is both compatible
with the elimination of tachyons from external legs, and also Lorentz invariant as is
shown in ref. [7].

\section{Canonical procedures}

For simplicity's sake we take a real scalar field obeying the equation
\begin{equation}
\label{eq1}
(\Box+m^2)\phi=j
\end{equation}
Our metric will be (diag =-, +, +, ..., +) in a d-dimensional space-time. So,
eq. (1) has the wrong sign of the mass term, with respect to the usual Klein-
Gordon equation. The Lagrangian is
\begin{equation}
\label{eq2}
{\cal L}=-\frac {1} {2}\partial_\mu\phi\partial^\mu\phi
+\frac {1} {2} m^2\phi^2
\end{equation}
giving rise to the Hamiltonian $(\Pi=\dot{\phi})$
\begin{equation}
\label{eq3}
{\cal H}=\frac {1} {2}\Pi^2+
\frac {1} {2}\left(\nabla\phi\right)^2-
\frac {1} {2} m^2\phi^2
\end{equation}
Canonical quantization leads to
\begin{equation}
\label{eq4}
\left[\Pi,\phi^{'}\right]=-i\delta(\vec{r}-\vec{r}^{'})
\end{equation}
\begin{equation}
\label{eq5}
\left[{\cal H},\phi\right]=-i\dot{\phi}
\end{equation}
\begin{equation}
\label{eq6}
\left[{\cal H},\Pi\right]=-i\ddot{\phi}
\end{equation}
Taking the Fourier transform of the scalar field, we can write $(k^2\geq m^2)$
\begin{equation}
\label{eq7}
\phi=\int\frac {dk} {\sqrt{2w}}
\left(a_k exp[-ikx]+a_k^{*} exp[ikx]\right)
\end{equation}
with $k_0=\sqrt{\vec{k}^2-m^2}$

Inserting (7) in eq. (3), we get for the total Hamiltonian
\begin{equation}
\label{eq8}
\phi=\int dk\frac {1} {2}w
\left(a_k^{*}a_k+a_ka_k^{*}\right)
\end{equation}
${\cal H}$ is similar to the Hamiltonian of a bradyonic particle. 
The commutation rules (4), (5) imply
\begin{equation}
\label{eq9}
[a_k,a_k^{*}]=\delta(\vec{k}-\vec{k}^{'})
\end{equation}
The operators $a_k$ and $a_k^{*}$ obey the usual commutation relations for the creation and destruction operators of particles.

\section{Spectrum and eigenfunctions}

We will now turn our attention to the eigenvalue problem for 
${\cal H}$ (eq. (8)) with the
commutation relations (9) [8, 9].

For each degree of freedom $\vec{k}(\vec{k}^2>m^2)$ we simply write
\begin{equation}
\label{eq10}
h=\frac {1} {2}w
\left(a^{*}a+aa^{*}\right)
\end{equation}
It follows that
\[[h,a^{*}]=wa^{*}\;\;\;;\;\;\;[h,a]=-wa\]
so that, for any eigenstate of $h$, with eigenvalue $E$,
\[h|E>=E|E>\]
we have
\begin{equation}
\label{eq11}
ha^{*}|E>=(E+w)a^{*}
\end{equation}
\begin{equation}
\label{eq12}
ha|E>=(E-w)a
\end{equation}
The operator $a^{*}$ (respectively $a$) increases 
(respectively decreases) the energy by the
amount $w$. These results follow from the form of $h$ 
and the commutation rules (10).

For a bradyonic particle
\begin{equation}
\label{eq13}
p^2+m^2=0
\end{equation}
and the set of all $p_\mu$, satisfying this equation is a two-sheeted hyperboloid. 
Each of the
two sheets is Lorentz invariant and can be characterized by the sign of $p_0$. The
requirement that the energy of a bradyonic particle shall be positive is then Lorentz
invariant and consequently a base for the representation of (10) should exist with the
property that one of its vectors (the vacuum $|0>$) must obey $a|0>=0$. From this
vector, the entire base can be constructed by successive applications of a creation
operator $a^{*}$. In this way we built up the Fock space for bradyons.

The situation for tachyons is different. The set of all $p$, obeying
\begin{equation}
\label{eq14}
p^2-m^2=0
\end{equation}
is now a one-sheeted hyperboloid, and no one of its proper parts is Lorentz invariant.

As a matter of fact, there is a symmetry with respect to the sign of the energy and the
base should exhibit this property. The spectrum should consist of a series of equally
spaced energy levels (separated by $w$) centred about the origin.

For a zero-energy eigenstate of $h$ we have
\[h|0>=0\therefore\left(a^{*}a+aa^{*}\right)|0>=0
\;\;\;\;\;(2aa^{*}-1)|0>=0\]
\begin{equation}
\label{eq15}
aa^{*}|0>=\frac {1} {2}|0>
\end{equation}
\begin{equation}
\label{eq16}
a^{*}a|0>=-\frac {1} {2}|0>
\end{equation}
We can now construct the whole of the Fock space for the tachyon by successive
applications of the increasing operator $a^{*}$ and the decreasing operator $a$:
\[a^{*m}|0>=\alpha_m|m>\;\;\;\;\;a^{m}|0>=\beta_m|-m>\]
\begin{equation}
\label{eq17}
aa^{*}|m>=\left(m+\frac {1} {2}\right)|m>
\end{equation}
\begin{equation}
\label{eq18}
a^{*}a|m>=\left(m-\frac {1} {2}\right)|m>
\end{equation}
\begin{equation}
\label{eq19}
h|m>=mw|m>
\end{equation}
\begin{equation}
\label{eq20}
h|-m>=-mw|-m>
\end{equation}
The label $m$ indicates the number of energy units $w$ contained in the state
$|m>$ and it
is a positive, negative or null integer. The energy spectrum has no upper or lower
bound. Obviously this means that the base will not have a positive-definite scalar
product, as $aa^{*}+a^{*}a$ is a positive operator in a Hilbert space. 
Our requirements
then lead us to the adoption of a base having indefinite metric. Such a type of metric
in quantum field theory has been studied in a systematic way in ref. [10]. Fortunately
we will not have to worry about the related formalism. As will be seen, free tachyons
will neither be able nor allowed to occupy those states.

Observe that, as $\vec{k}^2\rightarrow m^2$
($\vec{k}^2>m^2$), $w\rightarrow 0$
and the energy levels become more
and more dense.

Going back to the operators $a_k$ defined by (7), we find the following vacuum
expectation values:
\begin{equation}
\label{eq21}
<0|a_ka_k|0>=0=<0|a_k^{*}a_k^{*}|0>
\end{equation}
\begin{equation}
\label{eq22}
<0|a_ka_{k^{'}}^{*}|0>=\frac {1} {2}\delta(\vec{k}-\vec{k}^{'})
=-<0|a_{k^{'}}^{*}a_k|0>
\end{equation}
It is interesting to compare (22) with the vacuum expectation values of products of
bradyonic creation and destruction operators.

In a space with positive metric, Hamiltonian (10) (or that of eq. (4.5) of ref. [5])
must have a lower bound and this implies the existence of a state that is annihilated
by $a_k$ (see ref. [5J-V). In our case the operators $a_k^{*}$ and $a_k$ 
are merely increasing and
decreasing operators. They have a more symmetric role, as in (15) and (16) or (21)
and (22).

\section{Vacuum expectation values of products of field operators}

We first evaluate the vacuum expectation value for $phi$ given by (7). Taking into
account (21) and (22),
\[<0|\phi(x)\phi(y)|0>=
\int\frac{dk} {\sqrt{2w}}\int\frac{dk^{'}} {\sqrt{2w^{'}}}
(a_k exp[ikx]+a_k^{*}exp[-ikx])\cdot\]
\[\cdot (a_{k^{'}} exp[ik^{'}y]+a_{k^{'}}^{*}exp[-ik^{'}y])=\]
\[\int\frac{dk} {\sqrt{2w}}\int\frac{dk^{'}} {\sqrt{2w^{'}}}
\left\{<0|a_ka_{k^{'}}^{*}|0>exp[ikx - ik^{'} y] +\right.\]
\begin{equation}
\label{eq23}
\left.<0|[a_k^{*}a_{k^{'}}|0>exp [-ikx+ik^{'}y]\right\}=
\frac {1} {2}\int\frac {dk} {2w}
(exp [ik(x-y)]-exp [-ik(x-y)])
\end{equation}
Note that for bradyons we would only obtain the first exponential without the 1/2
factor. From (23) we can evaluate the vacuum expectation value of the chronological
product
\begin{equation}
\label{eq24}
<0|T\phi(x)\phi(y)|0>=
<0|\phi)(x)\phi(y)|0>\Theta(x_0-y_0)+
<0|\phi)(y)\phi(x)|0>\Theta(y_0-x_0)
\end{equation}
and, as (23) changes sign under the interchange of x and y,
\begin{equation}
\label{eq25}
<0|T\phi(x)\phi(y)|0>=\frac {1} {2}\int
\frac {dk} {2w}
(exp [ik(x-y)]-exp [-ik(x-y)]) Sg(x_0-y_0).
\end{equation}
For bradyons we would obtain
\[<0|T\phi^{br}(x)\phi^{br}(y)|0>=\]
\[\frac {1} {2}\int
\frac {dk} {2w}
(exp [ik(x-y)]\Theta(x_0-y_0)
-exp [-ik(x-y)])\Theta(y_0-x_0)=-i\Delta_F(x-y)\]
And it is easy to see that (25) can be rewritten as
\begin{equation}
\label{eq26}
<0|T\phi(x)\phi(y)|0>=-\frac {i} {2}[\Delta_F^T(x-y)+
\overline{\Delta_F^T(x-y)}]
\end{equation}
where $\Delta_F^T(x-y)$ is the truncated Feynman function defined in [6]. 
(The sphere $k^2-m^2$ is
suppressed.) We can also write (25) in the form
\begin{equation}
\label{eq27}
<0|T\phi(x)\phi(y)|0>=-\frac {i} {2}[\Delta_R^T(x-y)+
\Delta_A^T(x-y)]
\end{equation}
where $\Delta_R^T$ and $\Delta_A^T$ are the (truncated) advanced and retarded Green's functions,
respectively. Expressions (25), (26) and (27) show that their left-hand side coincides
with Cauchy's principal value at the poles $k_0 =w$ and $k_0=-w$
The Green's function just constructed can also be defined in terms of its Fourier
transform. When $k^2>m^2$ the expression $(k^2-m^2)^{-1}$ 
has two real poles at $k_0=\pm w$
and the $k_9$ integration for $\Delta_R^T$ (respectively $\Delta_A^T$) 
goes along the real axis leaving the
poles to the right (respectively left).

\newpage

\section{Discussion}

Canonical quantization of the bradyon field, as well as that for the tachyon field
(for $k^2>m^2$ ), leads to the usual creation and annihilation operators 
$a_k^{*}$ and $a_k$ which
determine the energy operator $h_k\sim\{a_k,a_k^{*} \}$ and obey the commutation rules
$[a_k,a_k^{*}] =\delta(\vec{k}-\vec{k}^{'}$. There are several bases to 
represent these relations and the
choice of the appropriate one must satisfy some physical requirements. For the case
of bradyons, the mass-shell relation $k^2+m^2=0=-k_0^2+\vec{k}^2+m^2$ defines a
two-sheeted hyperboloid, and each of the (Lorentz-invariant) sheets is characterized
by the sign of the energy. To seat in the positive (respectively negative) sheet, the
base must have a minimum (respectively a maximum) energy vector $|0>$, which is
nullified by $a_k$ (respectively $a_k^{*}$). The whole base can be constructed by means of
repeated applications of the increasing (respectively decreasing) operator $a_k^{*}$
(respectively $a_k$). This base only contains positive (respectively negative)
eigenvectors of the energy operator and constitutes an appropriate Lorentz-invariant
Fock space for bradyons.

For tachyons the situation is different. The mass hyperboloid $k^2-m^2=0$ is now
one-sheeted and none of its proper parts is Lorentz invariant. The base just described
is thus not suitable for this case. A base symmetric with respect to the sign of the
energy is required. Such a base can be constructed with the zero-energy eigenvector
$|0>$, which is annihilated by ${a_k,a_k^{*}}$, and acting on it by means of both the increasing
and decreasing operators. In this way we build up a base characterized by an integer
$|m>$ (for each degree of freedom) which can be positive, negative, or null, measuring
the number of energy units carried by the state. This is the Lorentz-invariant Fock
space (with indefinite metric) appropriate for the representation of tachyons. Once
this base is selected, it is a simple matter to evaluate the zero-energy expectation
value of the chronological product of field operators. The symmetry of the situation
suggests, and the calculation confirms, that the propagator is now the sum of the
retarded plus the advanced Green's functions. The tachyon propagator is then a
Cauchy's principal-value function. This result also allows us to understand why
tachyons cannot be found as free particles as Cauchy's propagator automatically
avoids the mass shell $\delta$-function and so no interaction can put a 
tachyon line on the
mass shell. The corresponding Fock space cannot be populated and tachyons do not
appear in asymptotic states.

\section*{Appendix}

Although in this paper we only wanted to evaluate the propagator, we think it
convenient to add some comments about the implementation of Lorentz
transformations.

The infinitesimal generators of the Poincard group can be written without
difficulties. They obey the usual algebra. For example the generators of Lorentz
boost are given by
\[M_{0i}^{(0)}=-\frac {i} {4}\int dk w(\partial_ia_k^{*}a_k+
a_k\partial_ia_k^{*}-a_k^{*}\partial_ia_k-
\partial_ia_ka_k^{*}\Theta(\vec{k}^2-m^2)\]
For a given spacelike momentum, all Lorentz transformations that do not change the
sign of the energy can be obtained by exponentiation. However, the implementation
of a Lorentz transformation that changes the sign of the energy needs special care. It
can be expressed as a product of a transformation that reduces the time component to
zero (without changing its sign) times another one that boosts the energy to its final
value. In between, to pass from $0^+$ to $0^-$(or viceversa), it is necessary to insert an
operator that takes the transposed conjugate. For example, the representative of a
Lorentz transformation that changes the sign of the energy must (as is well known)
change a creation operator into a destruction operator 
$a_k^{*}\longleftrightarrow a_{k^{'}}$. It also changes a
product $a_{k_1}^{*}a_{k_2}$ into $a_{k_2^{'}}^{*}a_{k_1^{'}}$.

\end{document}